\documentclass[a4paper,12pt,nonacm]{acmart}
\usepackage{graphicx}
\usepackage{tikz}
\usepackage{geometry}
\usepackage{xcolor}
\usepackage[T1]{fontenc}
\usepackage{tgbonum}
\usepackage{lmodern}
\usepackage{tgtermes}
\usepackage{tgpagella}
\usepackage{tgschola}
\usepackage{mathptmx}
\usepackage{utopia}
\usepackage{palatino}
\usepackage{bookman}
\usepackage{charter}
\usepackage{array}
\usepackage{longtable}
\usepackage{enumitem}
\setlist{nosep}
\usepackage{float}
\usepackage{algorithm}
\usepackage{algpseudocode}

\usepackage[resetlabels,labeled]{multibib}
\newcites{Memo}{Further Readings}

\definecolor{offwhite}{RGB}{220,220,220}

\newcommand{\papertitle}{A Public Dataset Tracking Social Media Discourse about the 2024 U.S. Presidential Election on Twitter/$\mathbb{X}$}

\newcommand{\paperauthors}{Ashwin Balasubramanian, Vito Zou, Hitesh Narayana, Christina You, Luca Luceri, Emilio Ferrara}
\newcommand{\paperaffiliation}{University of Southern California} 

\begin{document}

\begin{titlepage}
    \begin{tikzpicture}[remember picture, overlay]
        \node[anchor=north west, inner sep=0] at (current page.north west) {
            \includegraphics[width=\paperwidth, height=\paperheight, trim=275 375 275 375]{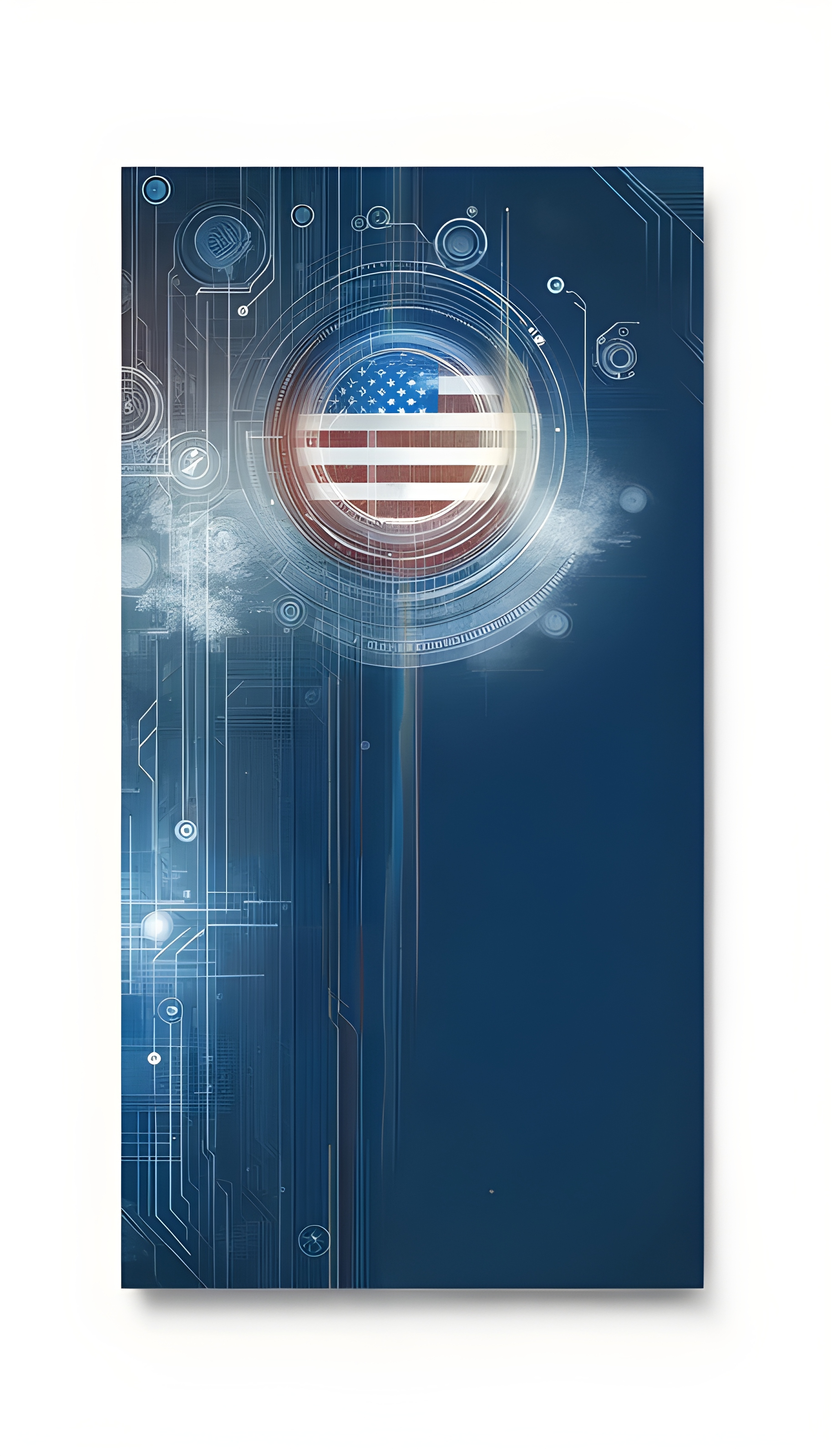}
        };
        \node[anchor=center, yshift=-9cm] at (current page.center) {
            \begin{minipage}{\textwidth}
                \raggedleft
                \color{offwhite}
                {\Huge \bfseries \fontfamily{qtm}\selectfont The 2024 Election Integrity Initiative }
                
                \vspace{1.5cm}
                
                {\LARGE \fontfamily{qtm}\selectfont \papertitle}
                
                \vspace{1.5cm}
                
                {\Large \fontfamily{qtm}\selectfont \paperauthors}
                
                \vspace{1cm}
                
                {\Large \fontfamily{qtm}\selectfont \paperaffiliation}
                
                \vfill
                
                {\Large \fontfamily{qtm}\selectfont HUMANS Lab -- Working Paper No. 2024.6}
            \end{minipage}
        };
    \end{tikzpicture}
\end{titlepage}

\noindent{\LARGE \fontfamily{qtm}\selectfont \papertitle}

\vspace{0.5cm}

\noindent{\large \fontfamily{qtm}\selectfont \paperauthors}

\noindent{\large \fontfamily{qtm}\selectfont \textit{\paperaffiliation}}

\section*{Abstract}
In this paper, we introduce the first release of a large-scale dataset capturing discourse on $\mathbb{X}$ (a.k.a., Twitter) related to the upcoming 2024 U.S. Presidential Election. Our dataset comprises 22 million publicly available posts on X.com, collected from May 1, 2024, to July 31, 2024, using a custom-built scraper, which we describe in detail. By employing targeted keywords linked to key political figures, events, and emerging issues, we aligned data collection with the election cycle to capture evolving public sentiment and the dynamics of political engagement on social media. This dataset offers researchers a robust foundation to investigate critical questions about the influence of social media in shaping political discourse, the propagation of election-related narratives, and the spread of misinformation. We also present a preliminary analysis that highlights prominent hashtags and keywords within the dataset, offering initial insights into the dominant themes and conversations occurring in the lead-up to the election.
Our dataset is available at: \url{https://github.com/sinking8/usc-x-24-us-election}

\section*{Introduction}
Social media has become an influential force in 21st-century politics globally. X.com (formerly Twitter) has been particularly significant in shaping political tensions and public opinion, offering researchers a valuable resource for studying the ideologies that are shared, the spread of misinformation, and the online campaigns supporting political movements and candidates [1, 5-7, 9]. Due to its open accessibility and vast reach, X.com plays a critical role in enabling political discourse.

With 611 million monthly active users,\footnote{https://www.statista.com/statistics/272014/global-social-networks-ranked-by-number-of-users/} $\mathbb{X}$ facilitates short-form, text-based interactions on a wide range of topics, making it a prime venue for public engagement in political discussions. It serves as a communication hub for prominent figures, including government officials and celebrities, who rely on $\mathbb{X}$’s platform to reach and influence millions. For instance, 2024 presidential candidates Donald Trump and Kamala Harris, who have garnered 92 million\footnote{https://socialblade.com/twitter/user/realdonaldtrump} and 21 million\footnote{https://socialblade.com/twitter/user/kamalaharris} followers respectively, use the platform to promote their campaigns and critique opponents. 
Such dynamics, where both political figures and their audiences play intertwined roles, make $\mathbb{X}$ a unique platform for analyzing political sentiments and trends in the context of the 2024 U.S. Presidential Election.

In this article, we introduce the $\mathbb{X}$ 2024 U.S. Presidential Election dataset, which contains posts and metadata capturing this dynamic environment. Through continuous data collection using targeted keywords and capturing significant events, our dataset provides researchers with an unprecedented view into how discourse on X.com is shaping and reflecting public opinion around the upcoming election. As data collection progresses, we aim to expand the dataset to capture emerging trends and pivotal shifts in political discourse, offering a valuable resource for future analyses.

\section*{Data Collection Framework}

\subsection*{X-Scraper Engine}
We developed a custom scraping engine, the X-Scraper, to systematically collect publicly-available data from X.com related to the 2024 U.S. Presidential Election. This scraper is designed to capture a range of post-specific details, including post type, content, media, and user metadata, allowing us to observe user behaviors and interactions over time. By employing this strategy, we aim to capture authentic and organically occurring patterns in the evolving political discourse [6].

In addition to individual posts, the scraper extracts user interface interactions, enabling insights into the broader content landscape and user engagement. This is especially significant for understanding the 2024 election cycle, as it allows us to analyze both content influence and reception among users. The X-Scraper employs flexible query structures with targeted keywords, timelines, and post types to ensure relevance and actionability in the collected data.

For technical details, including the algorithm structure and pseudo-code, see Appendix \S\ref{app:XScraperAlgorithm}.

\subsection*{Query Structure}
Our custom scraper uses a query structure similar to the X.com API, specifying parameters such as:
\begin{itemize}
    \item \textbf{since}: Start date for scraping posts.
    \item \textbf{until}: End date for scraping posts.
    \item \textbf{Keywords}: Targeted election-related keywords, structured in quotations. Multiple keywords are included in comma-separated format, with OR logic applied.
    \item \textbf{from}: Filters to include only posts by specific users.
    \item \textbf{filter}: Specifies post type (e.g., retweets, quotes, replies), used to collect retweets when general scraping did not yield them.
\end{itemize}

\noindent Example query string: 
\begin{quote}
\emph{("thedemocrats" OR "DNC" OR "Kamala Harris" OR "Dean Phillips" OR "williamson2024" OR "phillips2024" OR "Democratic party" OR "Republican party" OR "Third Party") until:2024-07-02\_00:00:00\_UTC since:2024-07-01\_00:00:00\_UTC}
\end{quote}
This query captures posts made between July 1 and July 2 in UTC that contain any of the specified keywords. For the list of tracked keywords, see the Appendix \S\ref{app:keywords}.

\subsection*{Methodology}
Our methodology involves dividing the timeline into smaller intervals, each queried with relevant keywords to account for specific events and discourse patterns. Regular sanity checks were performed to ensure continuous data coverage without gaps. 
 
As the scraper operates from the user interface, the dataset may not include all posts matching a query due to potential X.com’s interface restrictions. Nonetheless, continuous manual monitoring during the data collection revealed that X-Scraper captures the near totality of the publicly-available data that appear on the platform during the observation period. 
The scraper is implemented using a Chromium driver to navigate X.com efficiently, utilizing multiple personal accounts to streamline data collection within the platforms terms of service. Keywords were adjusted to capture timely and relevant events; for instance, “Assassination” was added following the attempt on Donald Trump on July 13, 2024, and “JD Vance” was included on July 15 when he was named Trump’s running mate, capturing a dynamic and contextually relevant subset of election discourse.

\section*{Exploratory Analysis}

\subsection*{Basic Summary of the Data}
The data schema, detailed in Table \ref{table:DataSchema} (cf., Appendix \ref{app:schema}), structures each entry with key fields that enhance understanding of tweet content and context, facilitating content's in-depth analysis:

\begin{itemize}
    \item \textbf{id}: A unique identifier for each tweet.
    \item \textbf{text}: The tweet's content, representing the user's message.
    \item \textbf{media}: Information on media attachments, enhancing engagement insights.
    \item \textbf{epoch}: A timestamp for tweet creation, allowing longitudinal analysis.
\end{itemize}
User interaction metrics such as \textbf{replyCount}, \textbf{retweetCount}, \textbf{likeCount}, and \textbf{quoteCount} quantify engagement, while fields like \textbf{conversationId} and \textbf{mentionedUsers} reveal conversation structures and social dynamics. This schema supports detailed exploration of user behavior and content influence in the 2024 election context, offering valuable insights into public discourse.

\subsection*{Top Keywords}

Table 1. highlights the ten most frequent keywords, with "Biden" and "Trump" leading. This prominence reflects their centrality in U.S. political discourse, underscoring their roles as focal points of debate and public opinion. The keyword "MAGA" (Make America Great Again) signals sustained engagement with the conservative base, reflecting how identity-based slogans maintain relevance in election rhetoric. The inclusion of "GOP," "Harris," and "conservative" illustrates the prominent alignment of discourse along party and ideological lines. Such insights are valuable for understanding how specific figures and terms drive public attention and conversation polarization.

\begin{table}[h]
    \centering
    \caption{Top 10 Keywords}
    \begin{tabular}{@{}lc@{}}
        \toprule
        \textbf{Keyword} & \textbf{Frequency} \\ \midrule
        Biden   & 11,070,686 \\
        Trump   & 5,730,951  \\
        MAGA    & 2,648,869  \\
        Joe     & 2,136,631  \\
        President & 1,671,385 \\
        GOP     & 1,560,280  \\
        Donald  & 1,534,099  \\
        Harris & 1,169,948 \\
        kamala &   984,206 \\
        conservative & 806,766 \\
        \bottomrule
    \end{tabular}
\end{table}

\subsection*{Top Hashtags}

Table 2 reveals the top 20 hashtags, illustrating major candidates and slogans such as $`\#maga`$, $`\#trump2024`$, and $`\#bidenharris2024`$. These hashtags underscore significant public engagement around key figures and movements, reflecting the electorate's ideological divides. The frequent appearance of $`\#voteblue2024`$ and $`\#project2025`$ further indicates active digital advocacy and long-term campaign strategies. Such recurring terms highlight how social media discourse aligns with, and at times shapes, campaign narratives and public sentiment during election cycles.

\begin{table}[h]
\centering
\caption{Top 20 Hashtags and Their Occurrences}
\begin{tabular}{@{}>{\centering\arraybackslash}p{5cm} >{\centering\arraybackslash}p{4cm}@{}}
    \textbf{Hashtag} & \textbf{Occurrences} \\
    \hline
    \#maga & 559,722 \\
    \#trump2024 & 341,792 \\
    \#trump & 170,923 \\
    \#bidenharris2024 & 166,669 \\
    \#biden & 165,277 \\
    \#donaldtrump & 40,780 \\
    \#biden2024 & 39,588 \\
    \#gop & 33,615 \\
    \#joebiden & 31,620 \\
    \#usa & 30,764 \\
    \#project2025 & 24,346 \\
    \#rnc & 22,907 \\
    \#foxnews & 20,813 \\
    \#democrats & 20,222 \\
    \#voteblue & 19,332 \\
    \#election2024 & 18,546 \\
    \#trump2024tosaveamerica & 18,455 \\
    \#fjb & 18,111 \\
    \#voteblue2024 & 17,477 \\
    \#kamalaharris & 17,436 \\
\end{tabular}
\end{table}

\begin{figure}[h]
    \centering
    \includegraphics[width=\textwidth]{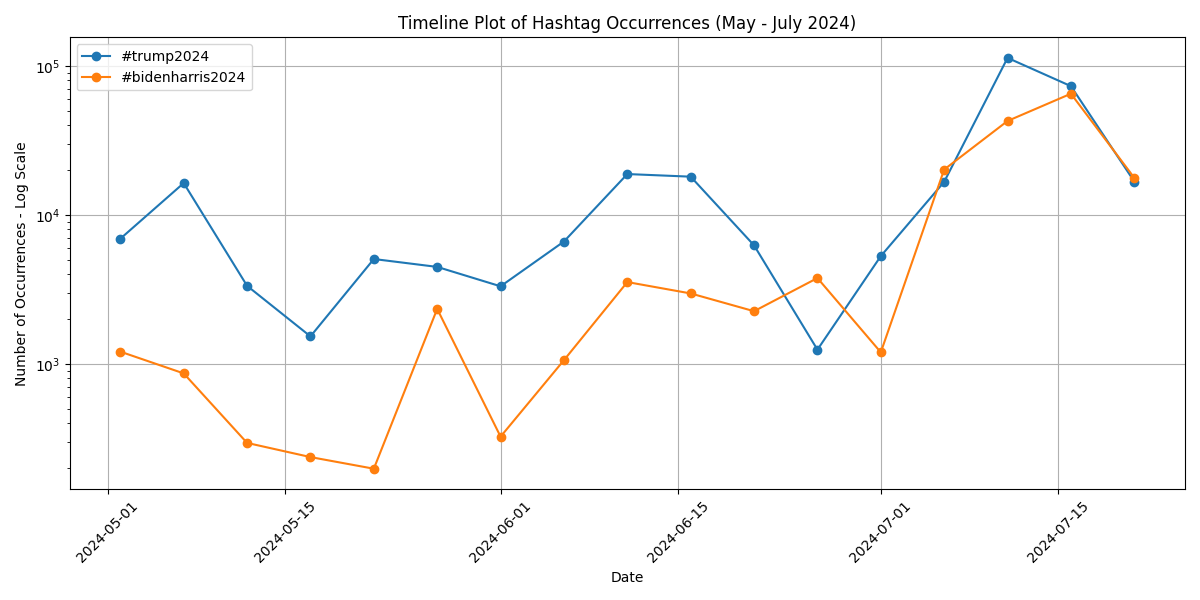}
    \caption{Timeline Plot of Hashtag Occurrences (May-July 2024)}
    \label{fig:my_image}
\end{figure}

Figure \ref{fig:my_image} shows the occurrences of hashtags $`\#trump2024`$ and $`\#bidenharris2024`$ over time, providing a timeline perspective that illustrates spikes in conversation, possibly due to specific campaign events or news cycles, thus offering insight into the immediate public reaction.

\subsection*{Top Social Media Domains}

Table 3 lists the top domains referenced, highlighting the popularity of YouTube and X.com for multimedia content sharing. News sites like Fox News and Breitbart underscore how major political narratives are supported and disseminated through recognized media outlets, shaping public discourse by providing prominent narratives to a wide audience. URL shorteners like dlvr.it also indicate strategic use of shortened links for spreading content efficiently across the platform.

\begin{table}[h]
    \centering
    \caption{Top 10 Domains and Their Frequencies}
    \begin{tabular}{c|c}
        \toprule
        \textbf{Domain} & \textbf{Count} \\
        \midrule
        youtu.be & 147,024 \\
        x.com & 125,817 \\
        dlvr.it & 79,870 \\
        www.youtube.com & 46,811 \\
        www.foxnews.com & 43,964 \\
        youtube.com & 40,954 \\
        trib.al & 36,267 \\
        www.msn.com & 29,724 \\
        l.smartnews.com & 27,091 \\
        www.breitbart.com & 26,443 \\
        \bottomrule
    \end{tabular}
\end{table}

\subsection*{Top 10 Mentions}

Table 4 displays the accounts most frequently mentioned, showing that public figures and entities play a central role in driving discourse on X. Mentions of candidates and key figures such as GOP, JoeBiden, TheDemocrats, and realDonaldTrump illustrate the polarized nature of political conversations, where users consistently engage with or critique prominent personalities. Elon Musk’s high mention frequency reflects his influence over the platform's direction and user perception, offering researchers an important perspective on how high-profile individuals impact public discourse.

\begin{table}[h]
    \centering
    \caption{Top 10 Mentions and Their Frequencies}
    \begin{tabular}{c|c}
        \toprule
        \textbf{Mention} & \textbf{Count} \\
        \midrule
        GOP & 423,427 \\
        JoeBiden & 358,955 \\
        TheDemocrats & 208,307 \\
        POTUS & 199,869 \\
        elonmusk & 146,573 \\
        realDonaldTrump & 139,868 \\
        YouTube & 127,044 \\
        harryjsisson & 108,825 \\
        KamalaHarris & 103,781 \\
        GuntherEagleman & 101,248 \\
        \bottomrule
    \end{tabular}
\end{table}

\section*{Conclusions}
This article introduces a comprehensive dataset of posts from $\mathbb{X}$ (formerly Twitter) related to the 2024 U.S. Presidential Election, collected between May 1, 2024, and July 31, 2024. Utilizing our custom-built X-Scraper Engine, which is designed to adapt to real-time events, we gathered a broad range of election-related content, including posts, metadata, and user information. 
This dataset provides a unique resource for researchers to analyze trends in public opinion, investigate the spread of misinformation, and examine the influence of key figures on $\mathbb{X}$. Given the polarized nature of the 2024 election cycle and the impact of social media on shaping public perceptions, this dataset has significant potential to help understanding how information and narratives are shared and propagated. Insights derived from this dataset could directly inform strategies to safeguard election integrity, mitigate misinformation, and assess the influence of prominent voices on digital political discourse.
While acknowledging limitations related to the representativeness of data from $\mathbb{X}$, we believe this dataset offers a useful sample of election-related content, providing a strong foundation for ongoing research into the dynamics between social media and political processes in the context of the 2024 U.S. Presidential Elections.

\subsection*{Limitations}
Our methodology presents certain limitations. Since the scraper collects data primarily through the user interface, it may not represent the entire timeline comprehensively. Additionally, updates to X.com’s interface occasionally introduced intermittent delays in scraping.
The use of keyword-based queries in $\mathbb{X}$’s search engine restricts the number of keywords per query, requiring multiple scraper runs to obtain a comprehensive subset of tweets. Another limitation is the fixed data retrieval rate of the scraper, capped at approximately 2,300 posts per hour per account due to $\mathbb{X}$’s terms of service. 

\subsection*{Future Work}
Future work will focus on continuous data collection throughout the election cycle, enabling a longitudinal analysis of political discourse and trends as the 2024 U.S. elections progress. Beyond mere volume, we plan to explore deeper patterns by examining verified users and suspected bots, to understand their roles in spreading information, shaping ideologies, and responding to real-time events. This will help identify the behaviors and strategies behind both authentic and inauthentic activity, providing a holistic view of platform usage during election times.

Additionally, we aim to incorporate data from complementary sources, such as cross-platform interactions and broader demographic metrics, to contextualize $\mathbb{X}$ discourse within the wider digital media landscape. This expanded approach will allow researchers to assess the platform’s unique role and interaction with other social media channels, leading to a richer understanding of election-related discourse across the online ecosystem.

\subsection*{Data Access}
Access to the dataset will be granted ensuring compliance with X.com’s policies. The repository will be updated consistently as scraping and processing progress, with periodic updates planned. Our dataset is available at: \url{https://github.com/sinking8/usc-x-24-us-election}

\section*{About the Team} \small
The 2024 Election Integrity Initiative is led by Emilio Ferrara and Luca Luceri and carried out by a collective of USC students and volunteers whose contributions are instrumental to enable these studies. The authors are indebted to Srilatha Dama and Zhengan Pao for their help in bootstrapping this data collection. The authors are also grateful to the following HUMANS Lab's members for their tireless efforts on this project: Ashwin Balasubramanian, Leonardo Blas, Charles 'Duke' Bickham, Keith Burghardt, Sneha Chawan, Vishal Reddy Chintham, Eun Cheol Choi, Priyanka Dey, Isabel Epistelomogi, Saborni Kundu, Grace Li, Richard Peng, Gabriela Pinto, Jinhu Qi, Ameen Qureshi, Tanishq Salkar, Kashish Atit Shah, Reuben Varghese, Christina You, Siyi Zhou.
\textbf{Previous memos}: Memo\citeMemo{memo1, memo2, memo3, memo4, memo5, memo7}

\renewcommand{\bibsection}{}

\clearpage
\section*{Appendix}

\subsection{X-Scraper Algorithm}\label{app:XScraperAlgorithm}

\begin{algorithm}
\caption{XScraper Algorithm}
\label{alg:XScraper}
\begin{algorithmic}[1]

\Function{XScraper}{outputfile, parameter\_data, cookies\_module}
    \State Initialize logger
    \State Set properties: outputfile, parameter\_data, cookies\_module, curr\_size, patience
\EndFunction

\Function{clear\_log\_file}{}
    \State Clear the log file
\EndFunction

\Function{get\_required\_time\_data}{epoch}
    \State Convert epoch to datetime and return month, day, year
\EndFunction

\Function{append\_csv\_to\_file}{filename, data}
    \State Append data to the specified CSV file
\EndFunction

\Function{patience\_check}{file\_path}
    \State Check if the current data size has changed and adjust patience
    \If {patience limit reached}
        \State Log and terminate scraping
    \EndIf
\EndFunction

\Function{append\_csv\_to\_file\_after\_pre\_processing}{file\_path, new\_data}
    \State Read existing data and append new data if available
    \State Reset patience if new data is found
\EndFunction

\Function{run}{cookies, search\_string, parameter\_data, outputfile}
    \State Launch browser and set up context
    \State Navigate to X.com and perform search
    \While {scraping}
        \State Intercept responses and update rate limiting information
        \If {patience check fails}
            \State Close browser and exit
        \EndIf
    \EndWhile
\EndFunction

\end{algorithmic}
\end{algorithm}
\clearpage

\subsection{Tracked Keywords} \label{app:keywords}

\begin{table}[h!] 
\centering
\caption{Tracked Keywords for 2024 Elections} 
\begin{minipage}[t]{0.45\textwidth}
    \footnotesize 
    \begin{tabular}{@{} >{\centering\arraybackslash}p{5cm} >{\centering\arraybackslash}p{4cm} @{}}
        \textbf{Keywords} & \textbf{Tracked Since} \\
        2024 Elections & 05/2024 \\
        2024 Presidential Election & 05/2024 \\
        Biden & 05/2024 \\
        Biden2024 & 05/2024 \\
        conservative & 05/2024 \\
        CPAC & 05/2024 \\
        Donald Trump & 05/2024 \\
        GOP & 05/2024 \\
        Joe Biden and Kamala Harris & 05/2024 \\
        Joe Biden & 05/2024 \\
        Joseph Biden & 05/2024 \\
        KAG & 05/2024 \\
        MAGA & 05/2024 \\
        Nikki Haley & 05/2024 \\
        RNC & 05/2024 \\
        Ron DeSantis & 05/2024 \\
        Snowballing & 05/2024 \\
        Trump2024 & 05/2024 \\
        trumpsupporters & 05/2024 \\
        trumptrain & 05/2024 \\
        US Elections & 05/2024 \\
        \\
        \\
    \end{tabular}
\end{minipage}%
\hspace{1cm} 
\begin{minipage}[t]{0.45\textwidth}
    \footnotesize 
    \begin{tabular}{@{} >{\centering\arraybackslash}p{5cm} >{\centering\arraybackslash}p{4cm} @{}}
        \textbf{Keywords} & \textbf{Tracked Since} \\
        thedemocrats & 05/2024 \\
        DNC & 05/2024 \\
        Kamala Harris & 05/2024 \\
        Marianne Williamson & 05/2024 \\
        Dean Phillips & 05/2024 \\
        williamson2024 & 05/2024 \\
        phillips2024 & 05/2024 \\
        Democratic party & 05/2024 \\
        Republican party & 05/2024 \\
        Third Party & 05/2024 \\
        Green Party & 05/2024 \\
        Independent Party & 05/2024 \\
        No Labels & 05/2024 \\
        RFK Jr & 05/2024 \\
        Robert F. Kennedy Jr. & 05/2024 \\
        Jill Stein & 05/2024 \\
        Cornel West & 05/2024 \\
        ultramaga & 05/2024 \\
        voteblue2024 & 05/2024 \\
        letsgobrandon & 05/2024 \\
        bidenharris2024 & 05/2024 \\
        makeamericagreatagain & 05/2024 \\
        Vivek Ramaswamy & 05/2024 \\
    \end{tabular}
\end{minipage}
\end{table}

\clearpage

\subsection{Data Schema} \label{app:schema}

\begin{table}[h!] 
\centering
\caption{Description of Data Schema} 
\label{table:DataSchema}
\begin{longtable}{|p{5cm}|p{3cm}|p{8cm}|}
\hline
\textbf{Field Name} & \textbf{Data Type} & \textbf{Description} \\
\hline
\endfirsthead
\hline
\textbf{Field Name} & \textbf{Data Type} & \textbf{Description} \\
\hline
\endhead
id & object & Unique identifier for each entry. \\
\hline
text & object & Text content of the tweet. \\
\hline
url & object & URL associated with the tweet or content. \\
\hline
epoch & object & Epoch timestamp when the tweet was created. \\
\hline
media & object & Media content included in the tweet (images, videos, etc.). \\
\hline
retweetedTweet & object & Content of the retweeted tweet, if applicable. \\
\hline
retweetedTweetID & object & ID of the retweeted tweet. \\
\hline
retweetedUserID & object & ID of the user who originally tweeted the retweeted content. \\
\hline
id\_str & object & ID of the tweet as a string (alternative format). \\
\hline
lang & object & Language of the tweet content. \\
\hline
rawContent & object & Raw unprocessed text of the tweet. \\
\hline
replyCount & object & Number of replies to the tweet. \\
\hline
retweetCount & object & Number of retweets. \\
\hline
likeCount & object & Number of likes. \\
\hline
quoteCount & object & Number of quotes. \\
\hline
conversationId & object & ID of the conversation the tweet is part of. \\
\hline
conversationIdStr & object & Conversation ID as a string. \\
\hline
hashtags & object & Hashtags included in the tweet. \\
\hline
mentionedUsers & object & Users mentioned in the tweet. \\
\hline
links & object & External links included in the tweet. \\
\hline
viewCount & object & View count of the tweet. \\
\hline
quotedTweet & object & Content of the quoted tweet, if applicable. \\
\hline
in\_reply\_to\_screen\_name & object & Screen name of the user being replied to. \\
\hline
in\_reply\_to\_status\_id\_str & object & ID of the tweet being replied to as a string. \\
\hline
in\_reply\_to\_user\_id\_str & object & User ID of the user being replied to as a string. \\
\hline
location & object & Location information of the tweet or user. \\
\hline
cash\_app\_handle & object & Cash App handle mentioned in the tweet, if applicable. \\
\hline
user & object & User information or metadata. \\
\hline
date & object & Date of the tweet. \\
\hline
\_type & object & Type of tweet (e.g., original, reply, retweet). \\
\hline
epoch\_dt & datetime64[ns] & Date and time in datetime format derived from epoch. \\
\hline
user\_id & float64 & ID of the user as a float. \\
\hline
\end{longtable}
\end{table}

\end{document}